\documentclass[smallextended]{svjour3}       
\smartqed  
\usepackage{graphicx}
\usepackage{float}
\usepackage{subfig}
\usepackage{amsmath}
\usepackage{bm}
\usepackage{amsmath}
\usepackage{amssymb}
\usepackage{latexsym}
\usepackage{epsfig}
\usepackage{amsbsy}
\usepackage{array}
\usepackage{setspace}

\usepackage{upgreek}
\journalname{Plasmonics}
\begin{document}

\title{A spectrally tunable plasmonic photosensor with an ultrathin semiconductor region}



\author{Shuyuan Xiao \and Tao Wang \and Xiaoyun Jiang \and Boyun Wang \and Chen Xu}

\institute{Shuyuan Xiao \and Tao Wang($\boxtimes$) \and Xiaoyun Jiang
           \at Wuhan National Laboratory for Optoelectronics, Huazhong University of Science and Technology, Wuhan 430074, People's Republic of China \\
           \email{wangtao@hust.edu.cn}
           \and Boyun Wang
           \at School of Physics and Electronic-information Engineering, Hubei Engineering University, Xiaogan 432000, People's Republic of China \\
           \and Chen Xu
           \at Department of Physics, New Mexico State University, Las Cruces 88001, United State of America}

\date{Received: date / Accepted: date}

\maketitle

\begin{abstract}
Surface plasmon resonance (SPR) has been widely utilized to improve the absorption performance in the photosensors. Graphene has emerged as a promising plasmonic material, which supports tunable SPR and shows significant flexibility over metals. In this letter, a hybrid photosensor based on the integration of periodic cross-shaped graphene arrays with an ultrathin light-absorbing semiconductor is proposed. A tenfold absorption enhancement over a large range of the incidence angle for both light polarizations as well as a considerably high photogeneration rate ($\sim10^{37}$) is demonstrated at the resonance. Compared with traditional metal-based plasmon-enhanced photosensors, the absorption enhancement here can be expediently tuned with manipulating the Fermi energy of graphene. The proposed photosensor can amplify the photoresponse to the incidence light at the selected wavelength and thus be utilized in photosensing with high efficiency and tunable spectral selectivity in the mid-infrared (mid-IR) and terahertz (THz) regime.
\end{abstract}

\section{Introduction}\label{sec1}
In recent years, surface plasmon resonance (SPR) has been introduced as an effective approach to reduce the thickness of the light-absorbing semiconductors in the photosensors, while maintaining a sufficient light absorption, leading to highly efficient photogeneration of charge carriers and current collection\cite{le2009plasmon,knight2011photodetection,choi2013versatile,wu2015surface,luo2016surface}. The basic physics lies in that the collective electronic excitations at metal/dielectric interface trap the incidence light in the near field, and induce the effects of electromagnetic field enhancement and light energy concentration. In the past, a variety of metal-based plasmonic structures have been proposed to integrate with the light-absorbing semiconductors to improve their absorption performance\cite{song2013great,cai2015tunable,xiong2015ultrabroadband,guo2016ultra,wang2016design,yi2017multiple,yu2017metamaterial}. Nevertheless, once the metal structures are fabricated, the resonance wavelength and thus the operation range of these hybrid photosensors will be unchangeable, which greatly hamper the flexible applications in practice.

Graphene, a monolayer of carbon atoms arranged in plane with a honeycomb lattice, behaves like metals when interacting with the incidence light and supports SPR for active material applications in the mid-infrared (mid-IR) and terahertz (THz) regime\cite{grigorenko2012graphene,abajo2014graphene,li2014tunable,he2016further}. Furthermore, the continuously tunable surface conductivity of graphene with manipulating its Fermi energy enables actively tunable resonance\cite{li2013investigation,lin2015combined,linder2016graphene,fu2016tunable,yan2016high,xiao2017strong}, which can be utilized to enhance the light absorption at the selected wavelength and thus amplify the photoresponse to the incidence light with tunable spectral selectivity\cite{fang2012graphene}. So far tunable absorption enhancement with graphene SPR has been extensively studied, however, most of previous studies concentrated on the absorption enhancement in graphene itself while the utilization of graphene SPR to trap light and enhance the light absorption in other light-absorbing semiconductors is very limited.

To explore this possibility, in this letter, a hybrid photosensor based on the integration of periodic cross-shaped graphene arrays with an ultrathin light-absorbing semiconductor region is numerically studied. The simulation results reveal that the excitations of SPR in the cross-shaped graphene arrays trap a substantial part of the incidence light in the near field and lead to a tenfold absorption enhancement in the surrounding light-absorbing semiconductor. With manipulating the Fermi energy of graphene, the absorption enhancement can be expediently tuned. Compared with the metal-based devices, our proposed photosensor can provide a comparable photogeneration rate and operate over a large spectral range.

\section{The geometric structure and numerical model}\label{sec2}
\begin{figure}[htbp]
\centering
\includegraphics[scale=0.4]{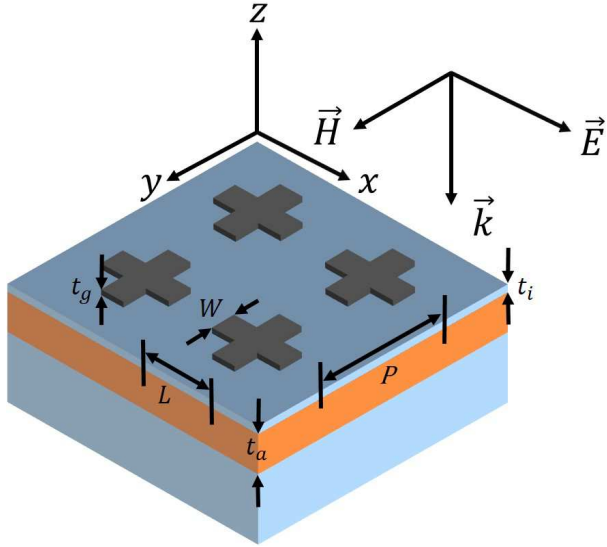}
\caption{\label{fig:1}The schematic geometry of our proposed hybrid periodic array. Each unit cell is composed of a cross-shaped graphene resonator on the top of the light-absorbing semiconductor separated by an insulating spacer.}
\end{figure}

The schematic geometry of our proposed photosensor is illustrated in FIG.~\ref{fig:1}. The unit cell is arranged in a periodic array with a lattice constant $P=400$ nm and composed of a cross-shaped graphene resonator on the top of the light-absorbing semiconductor separated by an insulating spacer. The length and the width of the cross-shaped graphene resonator are $L=240$ nm and $W=80$ nm, and the effective thickness is set as $t_g=1$ nm. The thicknesses of the insulating spacer and the light-absorbing semiconductor are $t_i=20$ nm and $t_a=100$ nm, and the substrate is assumed to be semi-infinite. Both the insulating spacer and the substrate are considered as lossless dielectrics with a real permittivity of $\varepsilon_d=1.96$. Comparable with some typical materials employed for photosensing in the mid-IR and THz regime, such as Hg$_{1-x}$Cd$_x$Te ternary alloy, the ultrathin light-absorbing semiconductor is modeled through a complex permitivity of $\varepsilon_a=\varepsilon^{'}+i\varepsilon^{''}$, where $\varepsilon^{'}=10.9$ and $\varepsilon^{''}$ is related to the absorption coefficient $\alpha=0.1$ $\upmu$m$^{-1}$ accounting for the losses\cite{rogalski2005hgcdte}.

Graphene is considered as an anisotropic material: isotropic in the plane of sheet and non-dispersive out of the plane, whose permitivity can be described with a diagonal tensor. The in-plane component is $\varepsilon_{xx}=\varepsilon_{yy}=2.5+i \sigma_g/(\varepsilon_0\omega t_g)$, and the out-of-plane component is $\varepsilon_{zz}=2.5$\cite{gao2012excitation,zeng2014high,xia2016excitation}, where $\varepsilon_0$ is the permittivity of vacuum, and $\omega$ is the angular frequency of the incidence light. $\sigma_g=i e^2E_F/[\pi\hbar^2(\omega+i/\tau)]$ is the intraband Drude-like surface conductivity of graphene within the random-phase approximation (RPA)\cite{zhang2014coherent,zhang2015towards}, where $e$ is the charge of an electron, $E_F$ is the Fermi energy of graphene, and $\hbar$ is the reduced Planck's constant. $\tau=\mu E_F/(e v_F^2)$ is the relaxation time, which depends on the electron mobility $\mu\approx10000$ cm$^2$/Vs, the Fermi energy $E_F$ and the Fermi velocity $v_F\approx10^6$ m/s. To manipulate the Fermi energy of graphene, similar to that used in Ref. \cite{ju2011graphene,hu2015broadly}, the top gate configuration is included (not shown in FIG.~\ref{fig:1}), and a 100 nm thick ion-gel layer is coated on the top of graphene, which is described by a non-dispersive permittivity $\varepsilon_{ig}=1.82$. The numerical simulations are further performed by using finite-difference time-domain (FDTD) method. The anti-symmetric and symmetric boundary conditions are respectively employed in the $x$ and $y$ directions throughout the calculations except when studying the angle polarization tolerance, and perfectly matched layer (PML) boundary conditions are utilized in the $z$ direction along the propagation of the incidence plane wave\cite{berenger1994perfectly,berenger2007perfectly}.

\section{Simulation results and discussions}\label{sec3}
\begin{figure}[htbp]
\centering
\includegraphics[scale=0.4]{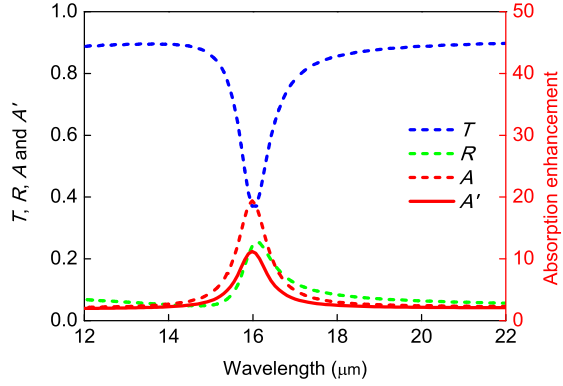}
\caption{\label{fig:2}The simulated transmission $T$, reflection $R$ and absorption $A$ as well as the absorption in the light-absorbing semiconductor $A'$ with the Fermi energy of graphene $E_F=0.6$ eV. The enhancement factor of absorption in the light-absorbing semiconductor is also shown compared to that in the impedance matched media.}
\end{figure}

Cross-shaped graphene resonator is a typical electric ring resonator, which strongly couples to the electric field of the incidence light. The resonance condition reads $2\pi n_{neff}L/\lambda_{res}=m\pi-\phi$\cite{ke2015plasmonic}, where $n_{eff}$ is the effective index depending on the geometric parameters of the structure and the Fermi energy of graphene, m=1,2,3, ... accounts for the order of the resonance, and $\phi$ is the phase change (modulus $\pi$) due to the reflection at the edges of the cross-arm. As the incidence wavelength is much larger than the length of the cross-shaped resonator, only the dipole resonance (m=1) is expected to survive. The simulated spectra with an initial Fermi energy of graphene $E_F=0.6$ eV are illustrated in FIG.~\ref{fig:2}. At the resonance around 16 $\upmu$m, the transmission is strongly suppressed and the absorption is significantly enhanced. The total absorption of the hybrid structure is $A=38.8\%$ and the absorption in the light-absorbing semiconductor is $A'=22.4\%$. Note that this ultrathin semiconductor is set to only 100 nm thick with the absorption coefficient $\alpha=0.1$ $\upmu$m$^{-1}$, corresponding to an extremely weak absorption of $2\%$ in the impedance matched media, a tenfold absorption enhancement is realized at the resonance.

\begin{figure}[htbp]
\centering
\subfloat[]{ \label{fig:3subfig:1} 
\includegraphics[scale=0.25]{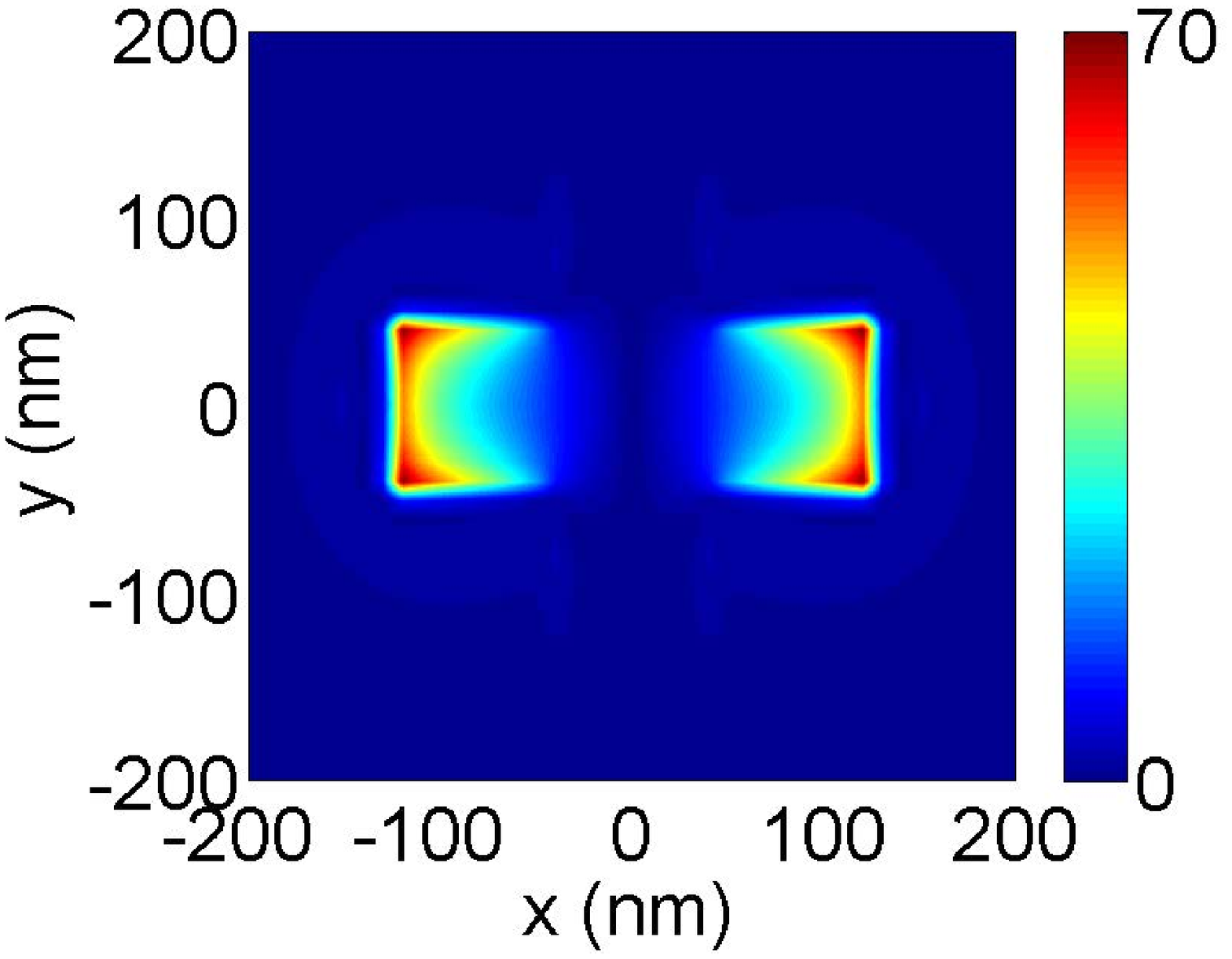}}
\subfloat[]{ \label{fig:3subfig:2} 
\includegraphics[scale=0.25]{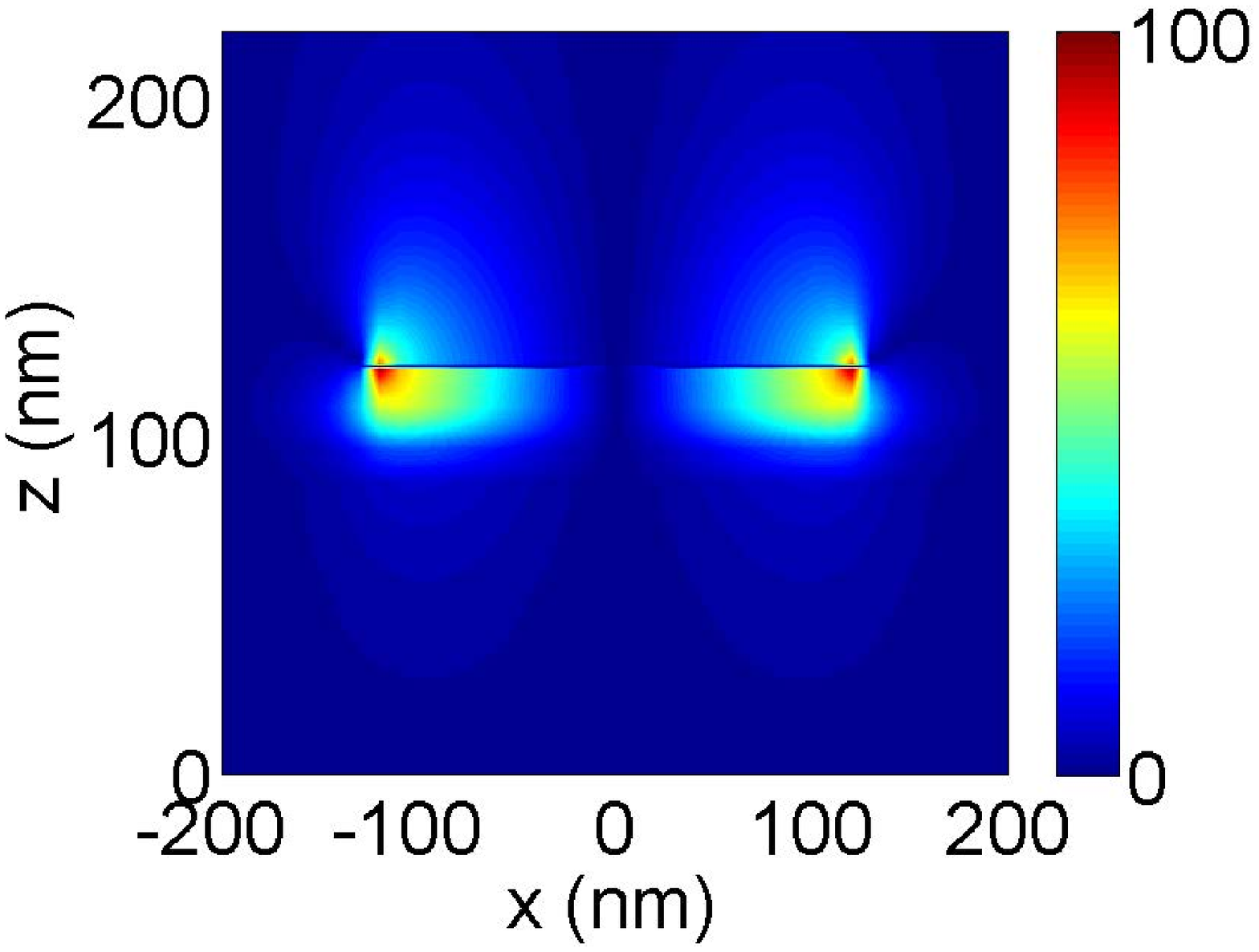}}\\
\subfloat[]{ \label{fig:3subfig:3} 
\includegraphics[scale=0.25]{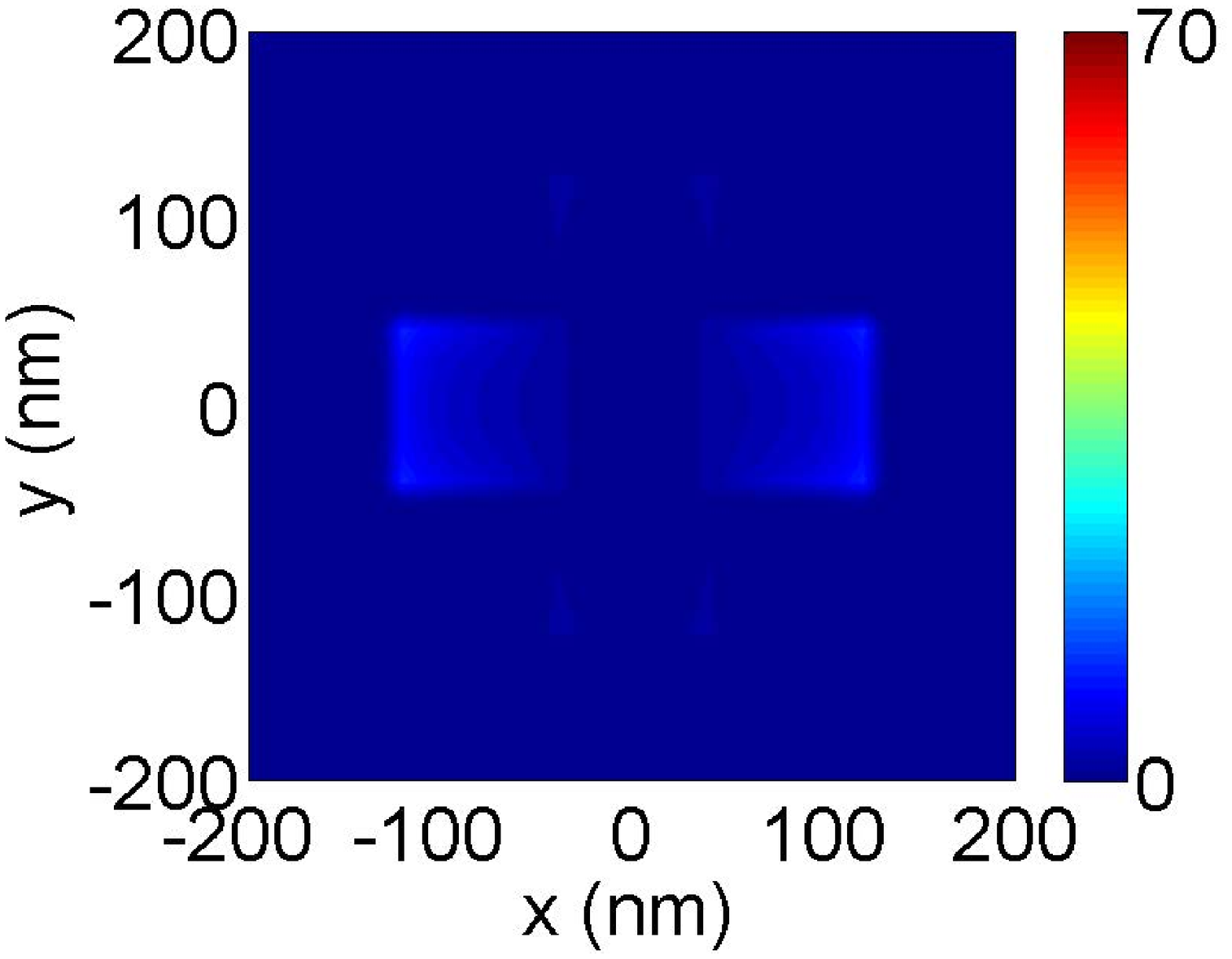}}
\subfloat[]{ \label{fig:3subfig:4} 
\includegraphics[scale=0.25]{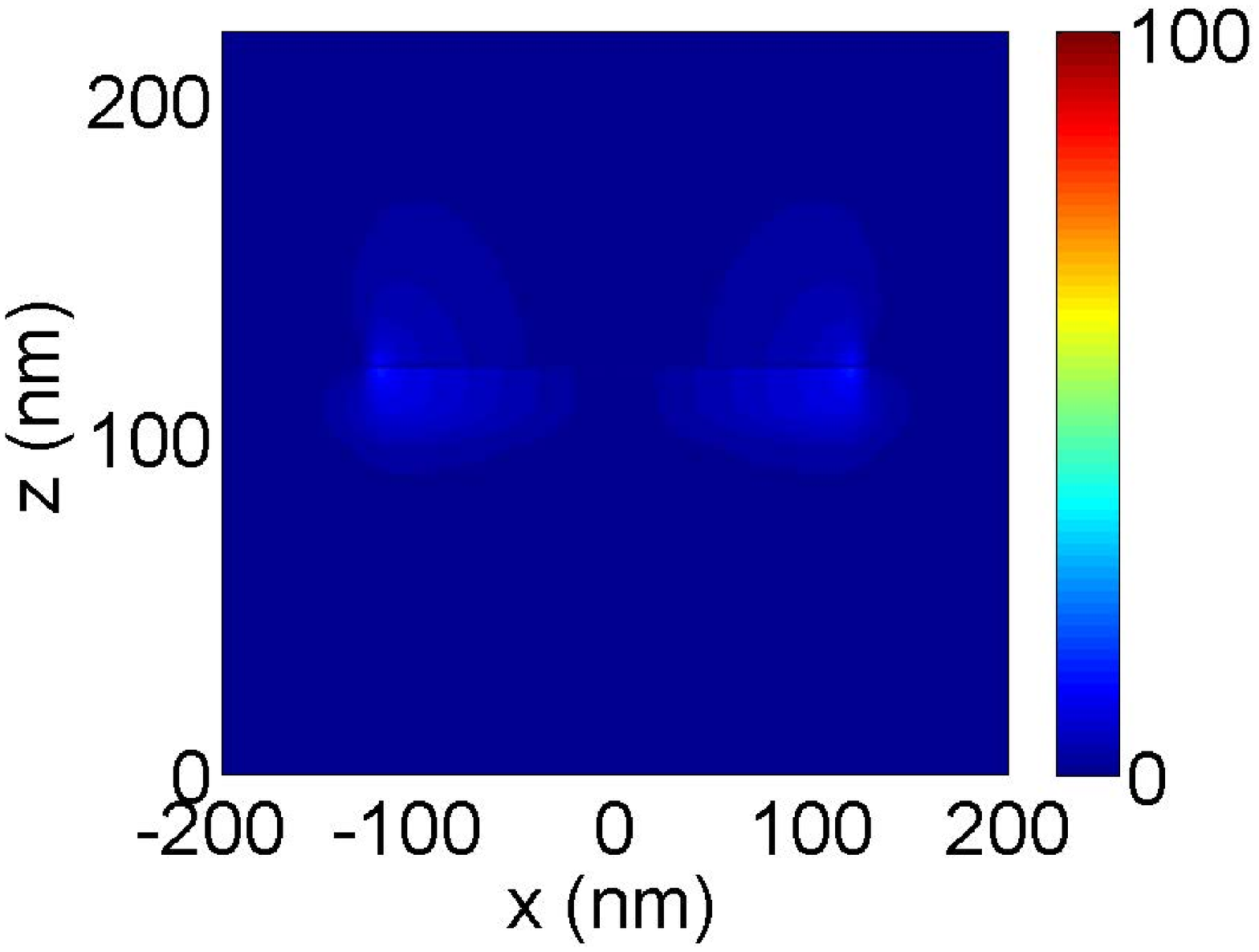}}
\caption{\label{fig:3}(a)-(d) The simulated electric field distributions with the Fermi energy of graphene $E_F=0.6$ eV. (a) $|E_z|$ and (b) $|E_y|$ at the resonance; (c) $|E_z|$ and (d) $|E_y|$ at 20 $\upmu$m.}
\end{figure}

In view of the theory of plasmonic metamaterials, the enhanced absorption performance in the light-absorbing semiconductor should be attributed to graphene SPR: the localized collective electronic excitations lead to light trapping and field enhancement surrounding the cross-shaped graphene resonator, and thus serve to enhance the light-matter interaction. Based on such consideration, the electric field distributions are essentially analyzed in FIG.~\ref{fig:3}. At the resonance, a strong enhancement of the $x$-$y$ plane electric field ($|E_z|$) is illustrated in FIG.~\ref{fig:3subfig:1} and it mostly concentrates on the edges of the cross-arm along the $x$ axis. This is a characteristic behavior of the electric dipole resonance in the cross-shaped resonator, which results from the accumulated charges at the edges. More notably, the $x$-$z$ cross plane electric field ($|E_y|$) in FIG.~\ref{fig:3subfig:2} clearly shows this plasmon-induced field enhancement extends to the surrounding region. Therefore, such strong resonance can effectively trap the light energy and provide sufficient time to dissipate it by the Ohmic losses in the light-absorbing semiconductor. In contrast, as illustrated in FIG.~\ref{fig:3subfig:3} and \ref{fig:3subfig:4}, there is nearly no field enhancement for enhanced absorption in the light-absorbing semiconductor at 20 $\upmu$m since this wavelength is far away from graphene SPR wavelength.

\begin{figure}[htbp]
\centering
\includegraphics[scale=0.4]{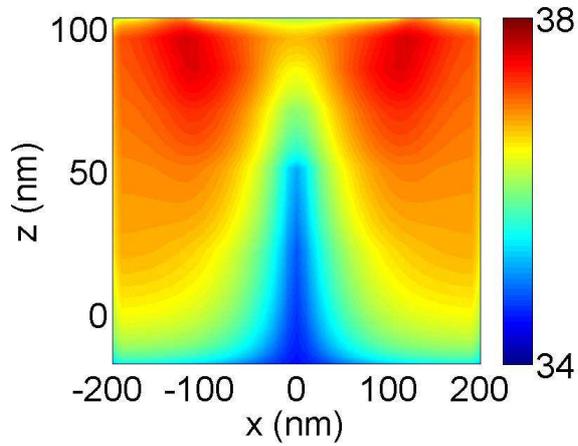}
\caption{\label{fig:4}The simulated photogeneration rate distribution (in base 10 logarithmic scale) with the Fermi energy of graphene $E_F=0.6$ eV at the resonance.}
\end{figure}

When dealing with a photosensing issue, it is necessary to further examine the photogeneration rate. In the light-absorbing semiconductor, the absorption per unit volume can be calculated from the divergence of the Poynting vector $Pabs=-0.5 real(\vec{\nabla}\cdot\vec{P})$, which however tends to be very sensitive to numerical calculations. Fortunately, it can be converted to a more numerically stable expression $Pabs=-0.5\omega|E|^2 imag(\varepsilon_a)$. And the number of absorbed photons per unit volume can then be calculated by dividing by the energy per photon $g=-0.5|E|^2 imag(\varepsilon_a)/\hbar$, in proportion to the square of the absolute value of the normalized local electric field $|E|^2$. Therefore, the strong electric field enhancement due to graphene SPR can provide a considerably high photogeneration rate. The photogeneration rate distribution (in base 10 logarithmic scale) at the resonance is plotted in FIG.~\ref{fig:4}. It can be seen that the photogeneration rate is largest ($\sim10^{37}$) neighboring the edges of the cross-arm along the $x$ axis and gradually becomes smaller toward the bottom of the light-absorbing semiconductor, showing a comparable performance with those in metal-based plasmon-enhanced photosensors\cite{le2009plasmon}.

\begin{figure}[htbp]
\centering
\includegraphics[scale=0.4]{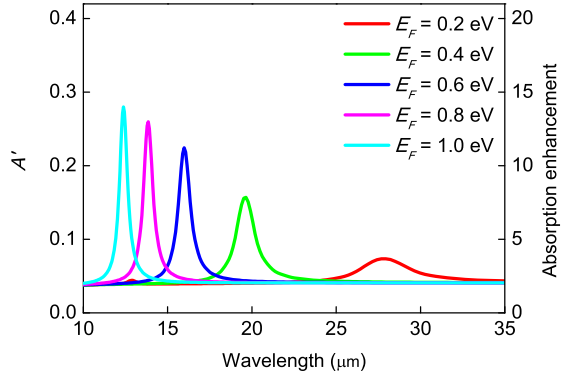}
\caption{\label{fig:5}The simulated absorption in the light-absorbing semiconductor $A'$ with the Fermi energy of graphene $E_F$ ranging from 0.2 to 1.0 eV. The enhancement factor of absorption in the absorbing layer is also shown compared to that in the impedance matched media.}
\end{figure}

The optical properties of graphene are largely determined by the Fermi energy. To demonstrate the spectral tunability of the proposed photosensor, the effects of the Fermi energy of graphene on the absorption enhancement are simulated in FIG.~\ref{fig:5}. When $E_F$ starts at 0.2 eV, the resonance wavelength locates at 27.9 $\upmu$m and the absorption in the light-absorbing semiconductor is $7.4\%$. As $E_F$ increases to 0.8 eV, the resonance shifts to 13.9 $\upmu$m and the absorption goes up to $26.0\%$. Finally, with a high Fermi energy of 1.0 eV, the absorption reaches $28.0\%$ at around 12.4 $\upmu$m. It can be seen that the resonance experiences a blue shift as the Fermi energy of graphene increases, and the absorption enhancement increases simultaneously. As mentioned above, the resonance wavelength can be described from the resonance condition $\lambda_{res}=2\pi n_{neff}L/(\pi-\phi)$, in positive proportion to the effective index $n_{eff}$. Note that $n_{eff}$ depends largely on the Fermi energy of graphene, and the higher $E_F$ suggests the smaller $n_{eff}$\cite{li2014sensitive,han2015dynamically,xiao2016tunable}. Hence the resonance wavelength will become shorter (corresponding to the blue shift of resonance). In the meanwhile, with the increase of the Fermi energy, the conductivity of graphene increases and and graphene SPR becomes less lossy, therefore the number of charge carriers contributing to the plasmonic resonance increases. As a result, the field enhancement with higher $E_F$ are stronger than that with lower $E_F$, which leads to a higher absorption enhancement in the light-absorbing semiconductor. As a whole, the tunability of the proposed photosensor with manipulating the Fermi energy of graphene lays the direct foundation for enabling tunable absorption enhancement in the light-absorbing semiconductor without changing the geometric parameters of the structure.

\begin{figure}[htbp]
\centering
\subfloat[]{ \label{fig:6subfig:1} 
\includegraphics[scale=0.25]{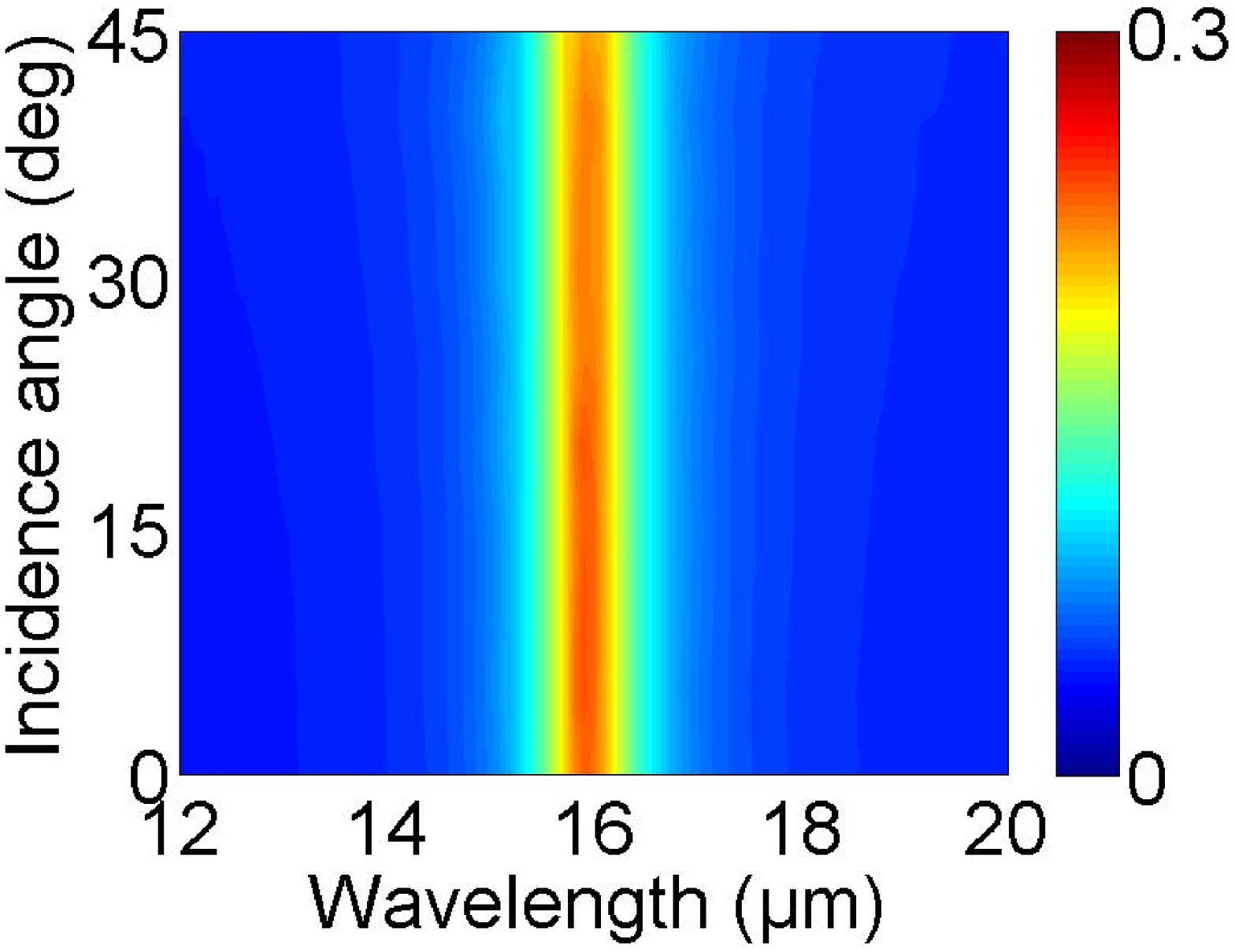}}
\subfloat[]{ \label{fig:6subfig:2} 
\includegraphics[scale=0.25]{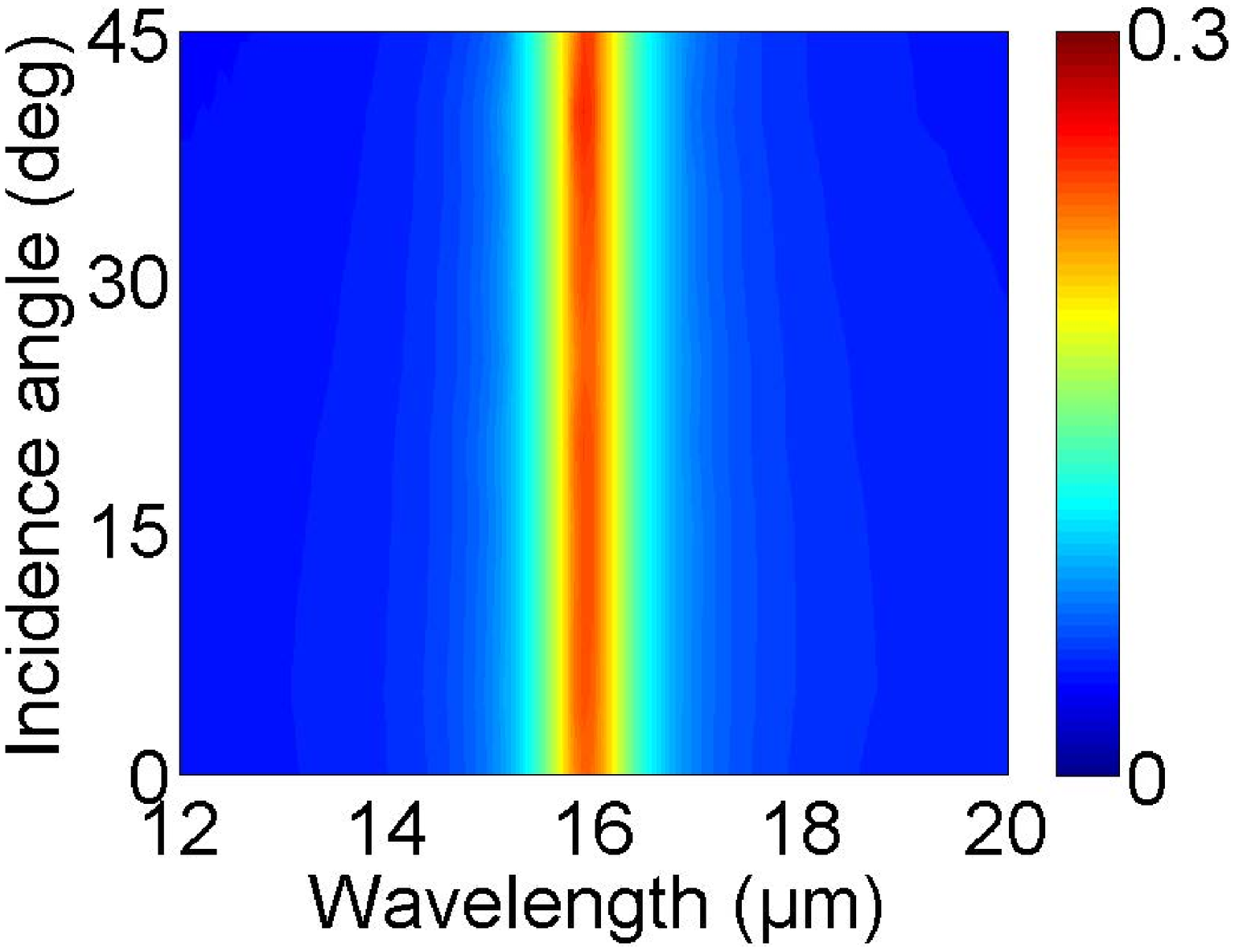}}
\caption{\label{fig:6}(a)-(b) The simulated angular dispersions of the absorption in the light-absorbing semiconductor with the Fermi energy of graphene $E_F=0.6$ eV for (a) TE and (b) TM configurations.}
\end{figure}

The dependences of the absorption in the light-absorbing semiconductor on the incidence angle and polarization are also investigated. As in FIG.~\ref{fig:6subfig:1} and \ref{fig:6subfig:2}, the resonance wavelength keeps exactly the same and the absorption also remains tenfold enhancement than in the impedance matched media over a large range of incidence angles [0$^{\circ}$, 45$^{\circ}$] for both TE and TM polarizations. The good operation angle polarization tolerance is mainly due to the following two reasons: firstly, the cross-shaped graphene resonator possesses the highly rotational symmetry; secondly, the absorption enhancement here results from the strongly localized surface plasmonic resonance. Consequently, the stable properties of absorption enhancement in the light-absorbing semiconductor is particularly desirable for realizing the robust photosensing platform.

\section{Conclusions}\label{sec4}
In conclusion, the strong possibility of improving the absorption performance in an ultrathin semiconductor region with graphene SPR is realized in our proposed photosensor. A tenfold absorption enhancement as well as a considerably high photogeneration rate ($\sim10^{37}$) is demonstrated at the resonance. Moreover, the tunability of graphene SPR with manipulating the Fermi energy enables actively tunable absorption enhancement in the surrounding light-absorbing semiconductor, which can amplify the photoresponse to the incidence light at the selected wavelength and thus be utilized in photosensing with high efficiency and tunable spectral selectivity. Although graphene SPR has mainly been observed in the mid-IR and THz regime, it has also been both theoretically suggested and experimentally demonstrated that graphene SPR can be pushed to much shorter wavelengths ($\sim2$ $\upmu$m)\cite{abajo2014graphene,wang2016experimental}, therefore the proposed photosensor together with its design principle can be applied to the near-IR regime.

\begin{acknowledgements}
The author Shuyuan Xiao (SYXIAO) expresses his deepest gratitude to his Ph.D. advisor Tao Wang for providing guidance during this project. SYXIAO would also like to thank Prof. Jianfa Zhang (National University of Defense Technology) for his guidance to the modeling of the light-absorbing semiconductor and Dr. Qi Lin (Hunan Univerisity) for beneficial discussion on graphene optical properties. This work is supported by the National Natural Science Foundation of China (Grant No. 61376055, 61006045 and 11647122), the Fundamental Research Funds for the Central Universities (HUST: 2016YXMS024) and the Project of Hubei Provincial Department of Education (Grant No. B2016178).
\end{acknowledgements}



\end{document}